\documentclass[preprint,12pt,3p]{elsarticle}

\usepackage{amssymb}
\usepackage{graphics}
\usepackage{epsfig}
\usepackage{subfigure}
\usepackage{epstopdf}

\expandafter\let\csname equation*\endcsname\relax 
\expandafter\let\csname endequation*\endcsname\relax 
\usepackage{amsmath}

\newcommand{\sech}{\mathop{\mathrm{sech}}\nolimits}

\usepackage{natbib}

\usepackage{dcolumn}% Align table columns on decimal point
\usepackage{bm}% bold math
\usepackage[colorlinks = true,
linkcolor = blue,
urlcolor  = blue,
citecolor = blue,
anchorcolor = blue]{hyperref}% add hypertext capabilities
\usepackage{amsmath}

\begin{document}

\begin{frontmatter}

\title{Variational approximations of soliton dynamics in the Ablowitz-Musslimani nonlinear Schr\"odinger equation}
%\tnotetext[label0]{This is only an example}
%
%
%\author[label1,label2]{Author One\corref{cor1}\fnref{label3}}
%\address[label1]{Address One}
%\address[label2]{Address Two\fnref{label4}}
%
%\cortext[cor1]{I am corresponding author}
%\fntext[label3]{I also want to inform about\ldots}
%\fntext[label4]{Small city}
%
%\ead{author.one@mail.com}
%\ead[url]{author-one-homepage.com}
%
%\author[label5]{Author Two}
%\address[label5]{Some University}
%\ead{author.two@mail.com}
%
%\author[label1,label5]{Author Three}
%\ead{author.three@mail.com}

\author[label2,label1]{Rahmi Rusin}
%\ead{rrusin@essex.ac.uk}

\author[label2,label3]{Rudy Kusdiantara}
%\ead{rkusdi@essex.ac.uk}

\author[label2]{Hadi Susanto\corref{cor1}}
\ead{hsusanto@essex.ac.uk}

\cortext[cor1]{Corresponding author at: Department of Mathematical Sciences, University of Essex, Wivenhoe Park, Colchester CO4 3SQ, United Kingdom.}
\address[label2]{Department of Mathematical Sciences, University of Essex, Wivenhoe Park, Colchester CO4 3SQ, United Kingdom}
\address[label1]{Department of Mathematics, Faculty of Mathematics and Natural Sciences, Universitas Indonesia, Ged D lt. 2 FMIPA Kampus UI Depok 16424, Indonesia}
\address[label3]{Theoretical Physics Laboratory, Theoretical High Energy Physics and Instrumentation Research Group, Faculty of Mathematics and Natural Sciences, Institut Teknologi Bandung, Bandung, 40132, Indonesia}
%\address[label4]{Centre of Mathematical Modelling and Simulation, Institut Teknologi Bandung, 1st Floor, Labtek III, Jl.\ Ganesha No.\ 10, Bandung, 40132, Indonesia}

\begin{abstract}
We study the integrable nonlocal nonlinear Schr\"odinger equation proposed by Ablowitz and Musslimani, that is considered as a particular example of equations with parity-time ($\mathcal{PT}$) symmetric self-induced potential. We consider dynamics (including collisions) of moving solitons. Analytically we develop a collective coordinate approach based on variational methods and examine its applicability in the system. We show numerically that a single moving soliton can pass the origin and decay or be trapped at the origin and blows up at a finite time. Using a standard soliton ansatz, the variational approximation can capture the dynamics well, including the finite-time blow up, even though the ansatz is relatively far from the actual blowing-up soliton solution. In the case of two solitons moving towards each other, we show that there can be a mass transfer between them, in addition to wave scattering. We also demonstrate that {defocusing} nonlinearity can support bright solitons. %We also propose and consider several variants of nonlocal equations, admitting nontrivial dynamics.
\end{abstract}

%\pacs{63.20.Pw, %localized modes
%	63.20.Ry, %anharmonic lattice modes
%}

\end{frontmatter}

% \linenumbers

\section{Introduction}

We consider the following Ablowitz-Musslimani nonlocal nonlinear Schr\"odinger (NLS) equation \cite{ablo13}
\begin{equation}
i\psi_t(x,t)+\frac12\psi_{xx}(x,t)+\sigma \psi^2(x,t)\psi^*(-x,t)=0,
\label{nnls}
\end{equation}
where $\psi\in\mathbb{C}$ is a complex valued function of the real variables $t\in\mathbb{R}^+$ and  $x\in\mathbb{R}$, and $^*$ denotes complex conjugation. The nonlinearity
coefficient is denoted by $\sigma$ where the value can be either $+1$ or $-1$, indicating the focusing or defocusing nonlinearity, respectively. The equation is integrable under the inverse scattering transform as it admits a linear (Lax) pair representation and possesses an infinite number of conservation laws \cite{ablo13,ablo16,ablo17,gerd17}. 

Equation \eqref{nnls} has the symmetry that it is invariant under the combined action of parity ($\mathcal{P}$) $x\to-x$ and time ($\mathcal{T}$) $t\to-t$, $\psi(x,t)\to\psi^*(x,t)$ operation, i.e., the governing equation is parity-time ($\mathcal{PT}$) symmetric.  The concept of $\mathcal{PT}$−symmetry itself was formulated by Bender and Boettcher \cite{bend98,bend99,bend07}, that has gained a lot of attention in the last decade \cite{such16,kono16}. It offers a `violation' to the standard postulate that the Schr\"odinger Hamiltonian operator be Dirac Hermitian and yet can have real eigenvalues up to a critical value of the complex potential parameter. Writing the nonlocal NLS equation \eqref{nnls} as a Hamiltonian with a complex potential $V(x,t)=\sigma\psi(x,t)\psi^*(-x,t)$, then the self-induced potential $V(x,t)$ can be straightforwardly shown to be $\mathcal{PT}$-symmetric. It is worth noting that although the introduction of $\mathcal{PT}$-symmetric models was motivated by the quantum-mechanical setting, the concept has been extended to a plethora of physical settings, see the reviews \cite{such16,kono16,gana18}. However, for the nonlocal NLS equation \eqref{nnls}, its physical relevance and experimental realisation are still a challenge. 

%The relevance of $\mathcal{PT}$-symmetry can be seen in many physical systems, such as   

The classical NLS equation  is recovered when the nonlocal nonlinear factor $\psi^*(-x,t)$ is replaced by $\psi^*(x,t)$. Because $\psi^*(-x,t)=\psi^*(x,t)$ when $\psi(x,t)$ is even, this tells us that even solutions to the classical NLS equation  will also be solutions to \eqref{nnls}. The nonlocal NLS equation  \eqref{nnls} can also be obtained from the defocusing NLS equation  under the variable transformation $x\to ix$ and $t\to-t$ \cite{yang18}. The equation admits periodically blowing-up one soliton solution \cite{ablo13}. However, initial conditions with rapidly decaying tails may not necessarily lead to blow-up \cite{ryba17}.

Various soliton solutions of the nonlocal NLS equation  \eqref{nnls} have been obtained, exploiting the integrability of the equations, such as dark and antidark solitons \cite{li15,feng18}, standing waves in terms of elliptic functions \cite{khar15}, and rational solutions both in the focusing and defocusing nonlinearity \cite{wen16,li16}. Note, however, that the study of soliton dynamics, especially interactions between solitons, has been done only using Darboux transformation \cite{li15, wen16, xu2017darboux,zhou2018darboux} or via a combination of Hirota's bilinear method and the Kadomtsev-Petviashvili (KP) hierarchy \cite{feng18}, which are limited to specific combinations of parameter values and initial conditions. A general study of interactions of many bright, dark and antidark solitons has been presented recently, where again the same transformation method was used \cite{priy18}. This paper addresses the problem of soliton dynamics numerically as well as semi-analytically using variational approximation (VA), which enables us to move beyond the limitation. %both for focusing and defocusing nonlinearity. 

VA has been a standard method in the study of solitons in NLS equation \cite{malo02}. However, it has not been applied and analysed for the nonlocal NLS equation  \eqref{nnls}. Here, we apply VA for the nonlocal NLS equation  \eqref{nnls} using the standard $\sech$ or Gaussian ansatz \cite{malo02}. We show that VA yields excellent agreement with the numerics for the dynamics of single localised waves. This includes the blow up one-soliton solution, which is quite interesting as our ansatz is relatively far from the actual solution. For soliton interactions, we observe that VA is only good prior to collisions at the origin as the centre of symmetry. 

In Sec.\ \ref{sec2}, we consider dynamics of single localised solutions. Gaussian wave packets that disperse when located far away from the origin and soliton solutions are also discussed. Interactions of two solitons are considered in  Sec.\ \ref{sec3}. In Sec.\ \ref{sec4}, we compare our analytical results in the previous two sections with numerical computations, where generally we obtain good agreement. We point out some limitations of our VA. We conclude our work with Sec.\ \ref{sec6}.

\section{Variational methods for a single soliton}
\label{sec2}

Consider a localised solution, i.e., a hump, with $|\psi(x,t)|\to0$ rapidly as $|x|\to\infty$. Let the centre of mass of the hump be located at $x=X$. When $|X|\to\infty$, one can observe that the nonlocal NLS equation  \eqref{nnls} becomes the nonrelativistic time-dependent linear Schr\"odinger equation studied in standard textbooks describing free particles moving in one dimension. As the linear Schr\"odinger equation has a localised in space solution describing wavepackets spreading in space as time evolves, a limiting solution of \eqref{nnls} in that case is given by \cite{ashb70}
\begin{equation}
\displaystyle
\psi=\sqrt{\frac{s}{\sqrt\pi\left(s^2+it\right)}}\exp\left[-\frac{1}{2}\frac{\left(x-p_0t-X\right)^2}{s^2+it}+i\frac{p_0(x-p_0t-X)}{2}\right],
\label{gw}
\end{equation}
where $s$ and $p_0$ are constants and $|X|\to\infty$.

Motivated by the dispersing wavepacket solution \eqref{gw}, to study dynamics of a Gaussian hump for the governing equation \eqref{nnls} using VA, we use the ansatz 
\begin{equation}
\psi=A\exp\left[iB\right]\exp\left[-C\left(x-X\right)^2\right]\exp\left[iD\left(x-X\right)^2+iE(x-X)\right],
\label{ans}
\end{equation}
where $A$ and $B$ account for the amplitude and the phase, $C$ is the decay rate, $D$ and $E$ are the chirp parameter that represents internal oscillations \cite{ande83,ande88} and the traveling velocity, respectively, and $X$ is the center of mass. The conventional travelling NLS soliton has a vanishing chirp $D=0$. 

In the following, we will allow the parameters to vary as a function of time, namely they become collective coordinates. The aim is to reduce the solution dynamics from being governed by the partial differential equation \eqref{nnls} into coupled ordinary differential equations that govern the dynamics of the collective variables  \cite{dawe13}. In doing so, we will use the variational method and as such, the collective coordinate approximation is also referred to as the VA.

The variational equations for the dynamics of the parameters are given by (see, e.g., \cite{rusi18})
\begin{eqnarray} \label{vareq}
Re\left[\int_{-\infty}^{\infty}\left(i\psi_t(x,t)+\psi_{xx}(x,t)+\sigma\psi(x,t)^2\psi^*(-x,t)\right)\frac{\partial }{\partial \square}\psi^*(x,t) \, \, dx \right]=0,
\end{eqnarray}
where $\square=A,B,C,D,E,X$. Explicit computations will yield the system of nonlinear differential equations 
\begin{subequations}
{\small \begin{align}
	\dot{A}&=\frac{\alpha_- \left(A^3 \sigma \left(\left(4 C^2 X^2-10 C-\beta_{-}^2\right) \sin \theta_--4 C \theta_- \cos \theta_-\right)-8 \sqrt{2} A C D \alpha_+\right)}{8 \sqrt{2} C},\\
	\dot{B}&=\frac{1}{8 \sqrt{2} C}\left(\alpha_- \left(A^2 \sigma \left(\left(-4 X^2 (C^2-D^2)+10 C+4 D E X-3 E^2\right) \cos \theta_-+4 C \theta_+ \sin \theta_-\right)\right.\right.  \nonumber\\
	&\quad \left. \left. -4 \sqrt{2} C \left(2 C-E^2\right) \alpha_+\right) \right) ,\\
	\dot{C}&=\frac{\alpha_- \left(A^2 \sigma \left(\left(4 C^2 X^2-2 C-\beta_{-}^2\right) \sin \theta_--4 C \theta_- \cos \theta_-\right)-8 \sqrt{2} C D \alpha_+\right)}{2 \sqrt{2}},\\
	\dot{D}&=\frac{\alpha_- \left(A^2 \sigma \left(\left(4 C^2 X^2-2 C-\beta_-^2\right) \cos \theta_-+4 C \theta_- \sin \theta_-\right)+4 \sqrt{2} (C^2-D^2) \alpha_+\right)}{2 \sqrt{2}},\\
	\dot{E}&=-\frac{\alpha_-A^2 \sigma  \left(\left(2 C^2 X+D \beta_-\right) \cos \theta_-+C (E-4 D X) \sin \theta_-\right)}{\sqrt{2} C},\\
	\dot{X}&=\frac{\alpha_-A^2 \sigma  (2 C X \sin \theta_--\beta_- \cos \theta_-)}{2 \sqrt{2} C}+E,
\end{align}}
\label{ODE_a}
\end{subequations}
with $\theta_{\pm}=X\beta_{\pm},\alpha_{\pm}=e^{\pm \frac{\beta_{-}^2}{4 C}\pm3 C X^2},\beta_{\pm}=E\pm2 D X.$

From the exact soliton solution in \cite{ablo13}, being trapped near the origin can make localised excitations blowing up. In that case, the ansatz \eqref{ans} will not be expected to describe a single soliton well, which can be improved by using the sech ansatz \cite{ande78,bond79}
\begin{equation}\label{ansatz}
\displaystyle \psi(x,t)= A\exp\left[iB\right]\sech[C(x-X)]\exp\left[iD\left(x-X\right)^2+iE(x-X)\right].
\end{equation}
Performing the same calculations, we obtain instead of \eqref{ODE_a} the set of equations up to $\mathcal{O}(X^2)$
\begin{subequations}
{\small \begin{align} 
	\dot{A}&=\frac{1}{4 \pi  C^3}\left(8 A \tilde{\sigma} X \tilde{E} \left(C^4-\pi ^2 E^2 \tilde{C}_{1,1}+\pi  E^2 \left(2 C E \tilde{T}-\pi  \tilde{C}_{1,1}\tilde{E}^2\right)\right) +3 (\pi -2) \pi  A C^3 D\right),\\
	\dot{B}&=\frac{1}{24 C^5}\left(\tilde{E}^3 \left(2 \pi ^2 C E\tilde{\sigma}H_2 \left(4 D X \tilde{C}_{5,11}-3 E \tilde{C}_{1,1}\right)-\eta_9+\eta_{-3}G_2 -3 \gamma_1+\gamma_3\right)\right),\\
	\dot{C}&=\frac{1}{6 \pi  C^2}\left(\tilde{E}^3 \left(4 \tilde{\sigma} X \left(3 C^4-2 \pi ^2 E^2 \tilde{C}_{1,1}\right) G_2-4 \tilde{\sigma}X \left(3 C^4+4 \pi ^2 E^2 \tilde{C}_{1,1}-6 \pi  C E^3 H_2\right)\right. \right. \nonumber \\
	&\quad \left. \left. +9 (\pi -2) \pi  C^3 D H_1^3\right)\right),\\
	\dot{D}&=-\frac{1}{\pi ^2 C^3}\left(2 \left(\pi  E \tilde{\sigma} \tilde{E} \left(\pi  C \tilde{T} K+2 E \left(2 \pi ^2 D X \tilde{C}_{1,1} \tilde{E}^2+C^2 \left(\left(6+\pi ^2\right) D X-E\right)\right. \right. \right. \right. \nonumber \\
	&\quad \left. \left. \left. \left. +\pi ^2 D E^2 X\right)\right)+2 \left(C^7-\pi ^2 C^3 D^2\right)\right)\right),\\
	\dot{E}&=-\frac{1}{15 C^5}\left(4 \pi  \tilde{\sigma}\tilde{E} \left(2 C^6 E X-5 C^4 \left(D-E^3 X\right)+C^2 E \left(10 \left(6+\pi ^2\right) D^2 X-15 D E+3 E^4 X\right)\right. \right.\nonumber\\
	&\quad \left. \left. +5 \pi  D \left(C \tilde{T} K+4 \pi  D E X \left(C^2+E^2\right) \tilde{E}^2\right)+10 \pi ^2 D^2 E^3 X\right)\right),\\
	\dot{X}&=E-\frac{1}{3 C^5}\left(\pi  \tilde{\sigma} \tilde{E}^3 \left(C^4+\pi  C H_2 K+3 C^2 E \left(2 \left(\pi ^2-2\right) D X+E\right)+G_2 \left(-C^4 \right. \right. \right. \nonumber \\
	&\quad \left. \left. \left. +C^2 E \left(2 \left(6+\pi ^2\right) D X-3 E\right)+2 \pi ^2 D E^3 X\right)+6 \pi ^2 D E^3 X\right)\right),
\end{align}}
\label{ODE2}
\end{subequations}
with \[\begin{aligned}\tilde{C}_{m,n}&=mC^2+nE^2,\,\tilde{\sigma}=A^2\sigma,\,
\tilde{E}=\text{csch}\left(\frac{\pi  E}{C}\right),\, \tilde{T}=\coth \left(\frac{\pi  E}{C}\right),\\
G_n&=\cosh \left(\frac{n \pi  E}{C}\right),\, H_n=\sinh \left(\frac{n \pi  E}{C}\right),\\
\gamma_n&=C^3 \left(C^2\tilde{C}_{1,3}-3 \pi ^2 D^2\right) H_n,\, K=C^2 (E-4 D X)+E^2 (E-12 D X),\\
\eta_n&= 4 \pi  \tilde{\sigma} \left(4 C^4 (E-D X) +C^2 E^2 \left(n \pi ^2 D X-30 D X+7 E\right)+n \pi ^2 D E^4 X\right).
\end{aligned}\]

%Make a remark on the difference with the standard NLS equation . 

Before we check the validity of the approximations \eqref{ODE_a} and \eqref{ODE2}, we will first derive VA when there are two solitons interacting. 

\section{Variational methods for soliton collisions}
\label{sec3}

Following the idea of Karpman and Solov'ev \cite{karp81}, we assume that the interacting solitons can be written as a superposition of two single solitons 
\begin{subequations}
	\begin{align}
	&\psi(x,t)=\sum_{j=1}^2\psi_j(x,t),\\
	&\psi_j=A_j\exp[iB_j]\sech[C_j(x-X_j)]\exp\left[iD_j\left(x-X_j\right)^2+iE_j(x-X_j)\right].
	\end{align}
	\label{ans2}
\end{subequations}
%where $j=1,2$.

From the discussion in Sec.\ \ref{sec2}, we know that placing both solitons on one side of the origin will only yield spreading of the wavepacket. We therefore in the following assume that 
\begin{align}
|\Delta X|=|X_2-X_1|\gg1,\,|X_a|=|X_2+X_1|\ll1,\,|E_a|=|E_2+E_1|\ll1,
\label{assm1}
\end{align}
i.e., the solitons are far away from the origin and moving towards each other with almost the same velocity. Additionally, 
\begin{align}
|\Delta\square|=|\square_2-\square_1|\ll1,\,\square=A,B,C,D,E
\label{assm2}
\end{align}
i.e., the solitons are almost identical. 

Substituting the ansatz \eqref{ans2} into the nonlocal NLS equation  \eqref{nnls} and computing the dominant terms in the vicinity of each soliton will yield the coupled equations (see, e.g., \cite{yang10} for derivations in the classical NLS equation )
\begin{eqnarray}
i{\psi_j}_t(x,t)+\frac12{\psi_j}_{xx}(x,t)+\sigma \psi_j^2(x,t)\psi_{3-j}^*(-x,t)=0,\,j=1,2.
\label{ceq}
\end{eqnarray}
Here, we neglect the terms $\psi_j^2\psi_{j}^*$, $2\psi_j\psi_{3-j}\psi_{3-j}^*$%$2\psi_{j}\psi_{j}^*\psi_{3-j}$, $\psi_{3-j}^2\psi_{j}^*$
, etc., (where functions with the conjugate are evaluated at $(-x,t)$) from the equation because they contribute to higher order corrections. Next,  to derive the dynamics of the parameters we apply the variational equation to \eqref{ceq} (see \eqref{vareq}) and use the assumptions \eqref{assm1} and \eqref{assm2}. 

Consider the variation with respect to, e.g., $A_1$, which will involve the integral of $\psi_1^2\psi_{2}^*(-x,t){\partial\psi_1^*(x,t)}/{\partial A_1}$, i.e., 
\begin{align}
A_1^2A_2\exp[2iB_1^2-iB_2]\sech^3[C_1z_1]\sech[C_2z_2]e^{i\Delta\phi},
\end{align}
with
\begin{align}
\Delta\phi=D_1z_1^2-D_2z_2^2+E_1z_1+E_2z_2,\,z_j=\left(x+(-1)^jX_j\right)^2.
\end{align}
In the vicinity of $x\approx X_j$ for the $j$th soliton, the assumptions \eqref{assm1} and \eqref{assm2} will lead us to the approximations (see \cite{yang10})
\begin{align*}
&\sech^3[C_1z_1]\sech[C_2z_2]\approx\sech^3[C_az_1]\sech[C_az_2],\\
&D_1z_1^2-D_2z_2^2=\frac12\Delta D\left(z_1^2+z_2^2\right)
+D_a\left(z_1^2-z_2^2\right)\approx D_a\left( X_1^2-X_2^2\right),\\
&E_1z_1+E_2z_2=E_a\left(z_1+z_2\right)+\frac12\Delta E\left(X_1+X_2\right)\approx\frac12\Delta E\left(X_1+X_2\right),
\end{align*}
where 
\[\square_a=\frac12\left(\square_1+\square_2\right).\]
Such asymptotic simplifications will lead us to the equations
\begin{subequations}
{\small \begin{align} 
	\dot{A_j}&=-\frac{A_j C_j \left(C_a \tilde{\sigma} \left(\left(\pi ^2-6\right) C_j^2+3 \pi ^2 C_a^2\right) (\beta_1+\sin (B_j-B_{3-j}))+3 \pi ^2 D_j \left(\chi_1+\chi_2\right)\right)}{6 \pi ^2 C_a^4},\\
	\dot{B_j}&=\frac{1}{12 C_a^4}\left( C_j \left(C_a^2 \tilde{\sigma} (\beta_3-(\beta_2 C_j+2 C_a (2 E_j-3 \beta_2)))+\chi_3-3E_j^2(\chi_2+2C_jC_a^2)\right. \right.  \nonumber \\
	& \quad \left. \left. +3 \pi ^2 D_j^2 (C_a-C_j)\right)\right) ,\\
	\dot{C_j}&=\frac{1}{3 \pi ^2 C_a^4}\left( C_j^2 \left(3 \pi ^2 D_j \left(-\chi_1-C_j C_a^2+C_a^3\right)- C_a \tilde{\sigma} \left(\left(\pi ^2-6\right) C_j^2+\pi ^2 C_a^2\right)\right. \right. \nonumber \\
	&\quad \left. \left. (\beta_1+\sin (B_j-B_{3-j}))\right)\right) ,\\
	\dot{D_j}&=\frac{C_j^3 \left(C_a^2 \tilde{\sigma} (2 C_a-C_j) (\cos (B_j-B_{3-j})-\beta_2)-\chi_4+E_j^2\chi_2+\pi ^2 D_j^2 (3 C_j-C_a)\right)}{\pi ^2 C_a^4},\\
	\dot{E_j}&=\frac{2 C_j \tilde{\sigma} (2 (C_j C_aX_a \cos (B_j-B_{3-j})+5 \beta_1 E_j)+5 \sin (B_j-B_{3-j}) (2 E_j-D_j X_a))}{15 C_a},\\
	\dot{X_j}&=-\frac{C_j (C_a \tilde{\sigma} X_a \sin (B_j-B_{3-j})+3 C_j E_j)}{3 C_a^2},
\end{align}}
\label{ODE3}\end{subequations}
with 
\begin{align*}
&\tilde{\sigma}=2A_1A_2\sigma,\,\beta_1=(2 D_a (X_j-X_{3-j})+\Delta E)X_a\cos (B_j-B_{3-j}),\\
&\beta_2=(2 D_a (X_j-X_{3-j})+\Delta E)X_a\sin (B_j-B_{3-j}),\,\beta_3=(C_j-6 C_a) \cos (B_j-B_{3-j}),\\
&\chi_1=-3 C_j^3+C_j^2 C_a,\,\chi_2=3 C_j C_a^2-3 C_a^3,\, \chi_3=C_j^3 C_a^2+3 C_j^2 C_a^3,\,\chi_4=C_a^2C_j^3+C_j^2C_a^3.\end{align*}

In the following section, we will examine the applicability of system \eqref{ODE3} numerically. 

\section{Numerical simulations}
\label{sec4}

\begin{figure}[tbhp!]
	\centering
	\subfigure[]{\includegraphics[scale=0.5]{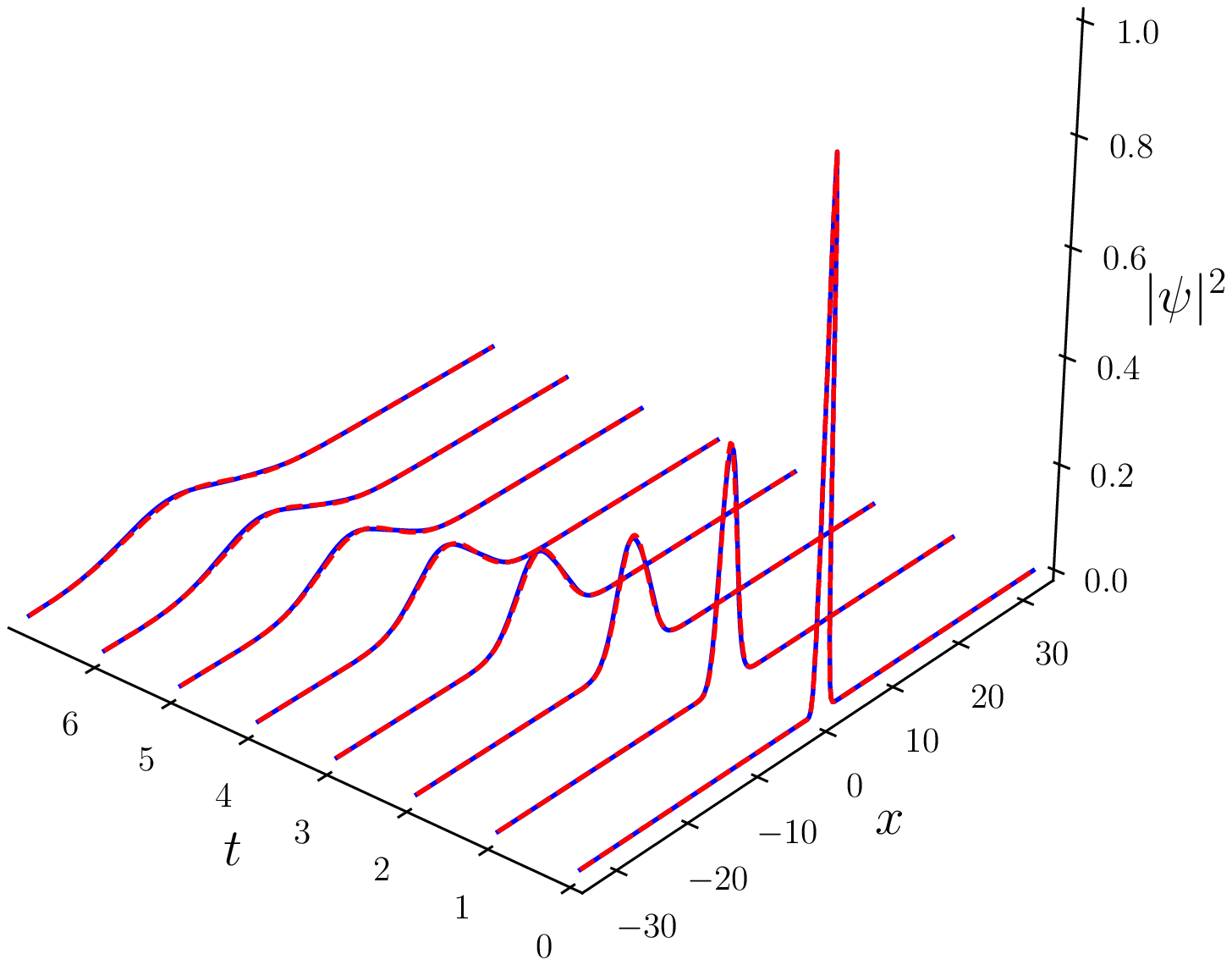}\label{subfig:FS_gauss_good1}}
	\subfigure[]{\includegraphics[scale=0.5]{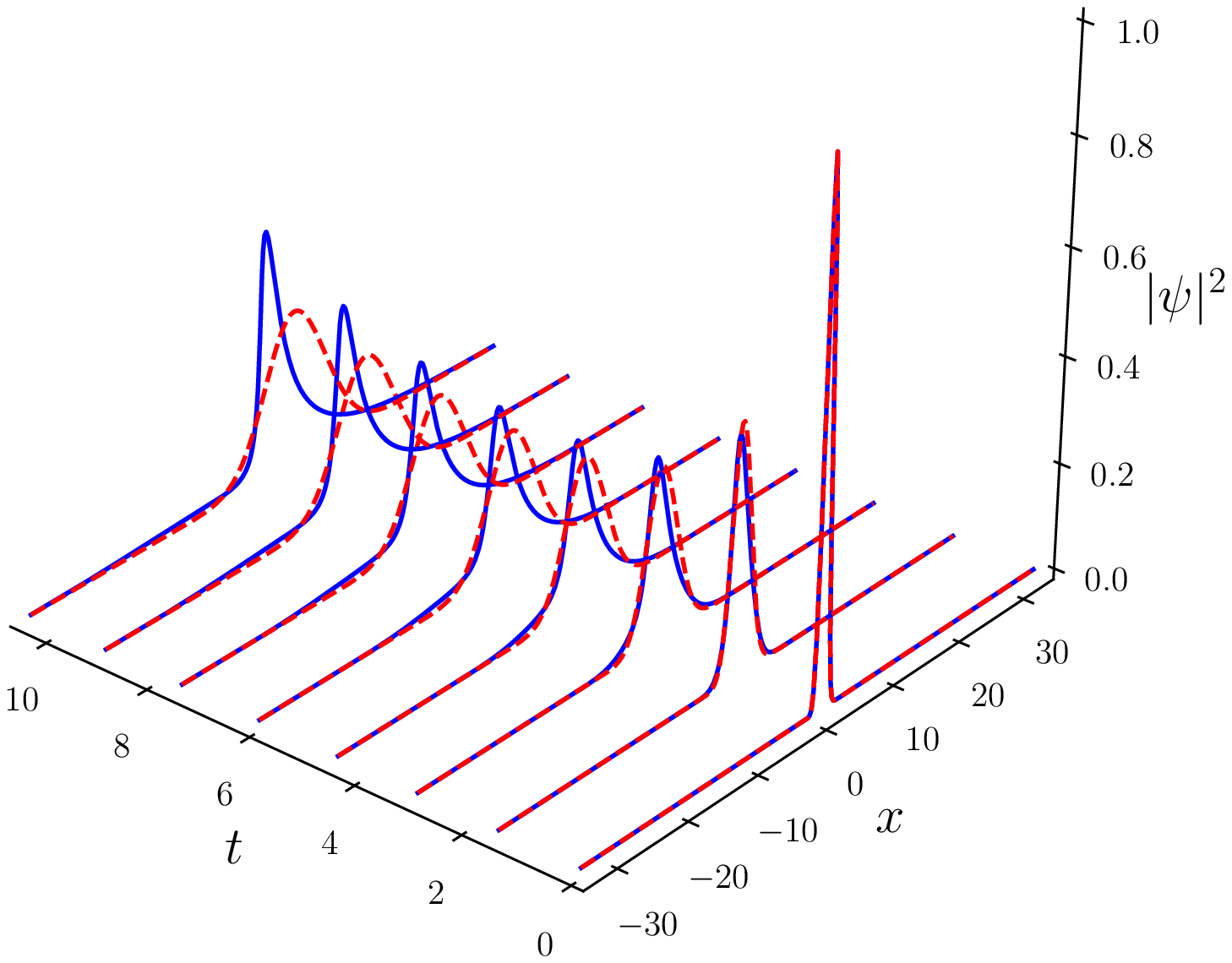}\label{subfig:FS_gauss_bad1}}
	\subfigure[]{\includegraphics[scale=0.5]{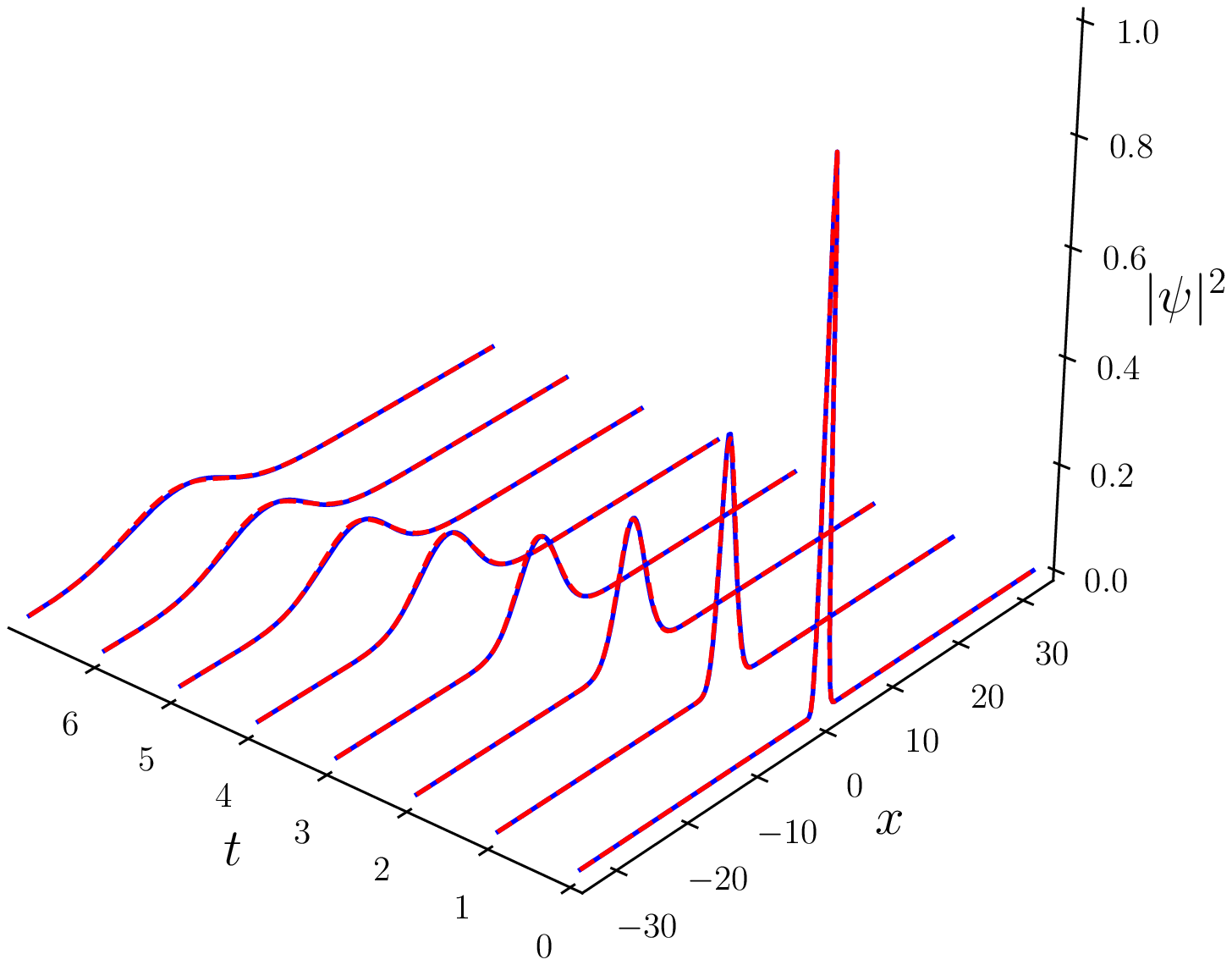}\label{subfig:DS_gauss_good1}}
	\subfigure[]{\includegraphics[scale=0.5]{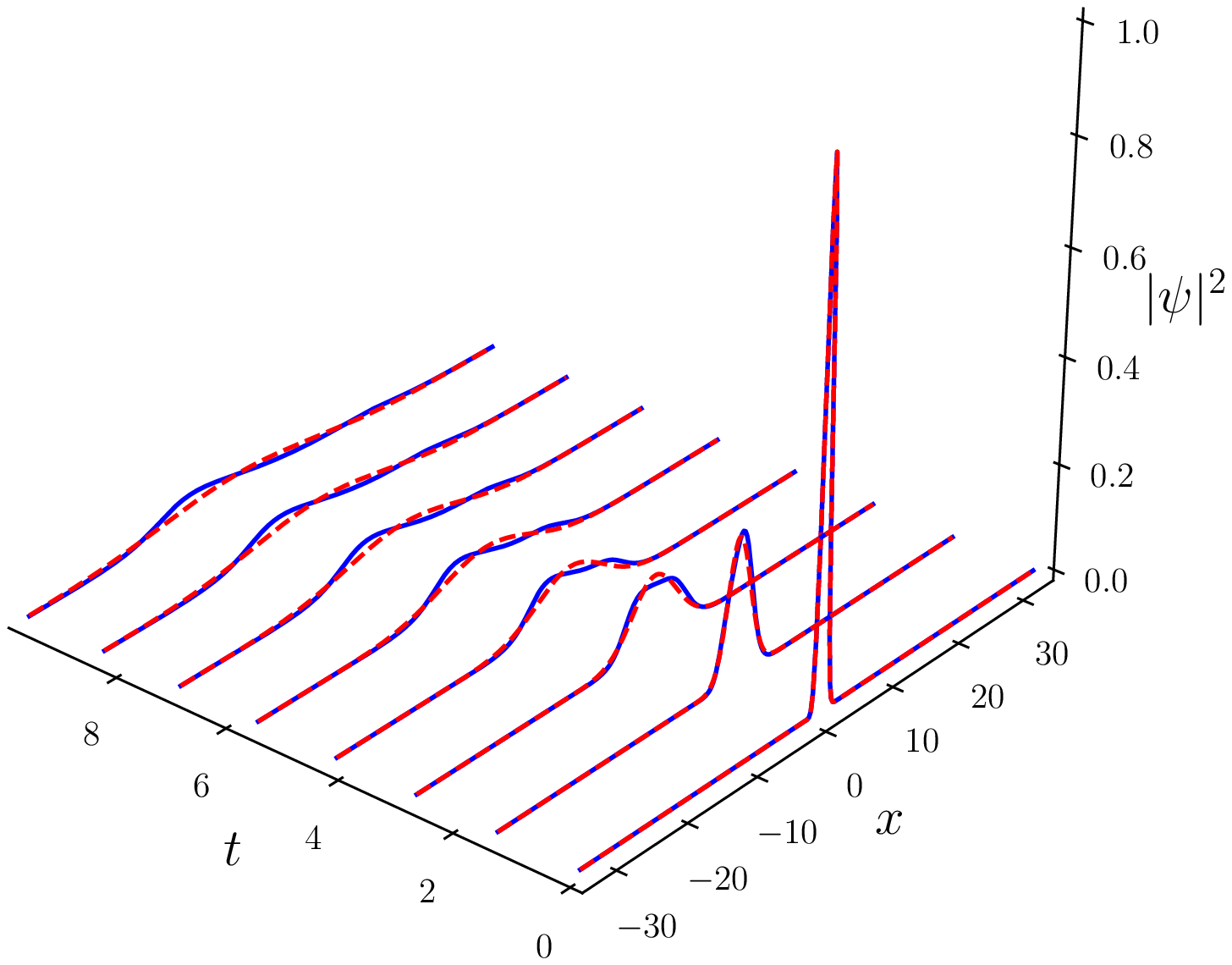}\label{subfig:DS_gauss_bad1}}
	\caption{Comparison of the travelling wave solutions of Eq.\ \eqref{nnls} between the numerical and variational solutions for the Gaussian initial condition \eqref{ans} with $A=1$, $B=0$, $C=1$, $D=0$, and $X=1$ in the focusing $\sigma=1$ (a,b) and defocusing $\sigma=-1$ (c,d) case. In (a,c), the initial velocity is $E=2$, while in (b,d), it is $E=0.1$. Blue solid and red dashed lines indicate the numerical and variational solutions, respectively. 
	}
	\label{fig:gauss1}
\end{figure}

We will compare the VAs \eqref{ODE_a}, \eqref{ODE2}, and \eqref{ODE3} with the soliton dynamics from the governing equation \eqref{nnls}. We numerically integrate \eqref{nnls} in time using the fourth-order Runge-Kutta method. The Laplacian is approximated by a pseudospectral differentiation matrix based on the Fourier series. Simulations below were carried out in the spatial interval $[L, L)$ with $L\geq 30$, and discrete stepsizes $\Delta x =0.1$ and $\Delta t = 0.005$ or smaller. By checking qualitative features of the evolution, we note that further decrease of $\Delta x$ and/or $\Delta t$ did not produce any conspicuous effect. The VAs \eqref{ODE_a}, \eqref{ODE2}, and \eqref{ODE3} are also integrated using the fourth-order Runge-Kutta method. The varying-in-time variables are then inserted back into the ansatz to obtain the spatial profile of the VA. 

First, we consider the passage of dispersive Gaussian wave packets. The initial condition is taken as \eqref{ans}. In Fig.\ \ref{fig:gauss1} we display the evolution of the incident Gaussian wave packet impinging onto the origin $x=0$ at two different values of initial velocities. Shown is the square absolute value of the field, $|\psi(x,t)|^2$. 

It is obtained that a large incoming velocity $E(0)$ yields rather excellent agreement between the numerics and the VAs. The type of nonlinearity is also not relevant for that case as panels (a) and (c) that are for focusing and defocusing nonlinearity, respectively, show quite similar time dynamics. This indicates that the system behaves as a linear one. 

When the incoming velocity is taken to be quite small, we obtain that the focusing case yields a trapped state that is followed by a blow-up. Unfortunately the Gaussian ansatz \eqref{ans} together with \eqref{ODE_a} can only capture the trapping. As for the defocusing case, there is no blow up. However, our VA also can only provide qualitative agreement. As shown in Fig.\ \ref{subfig:DS_gauss_bad1}, the wave function from the nonlocal NLS integration decays more rapidly than the VA. Moreover, there is a scattering and transmission process in the nonlocal NLS equation that cannot be possibly captured by our ansatz. 

\begin{figure}[tbhp!]
	\centering
	\subfigure[]{\includegraphics[scale=0.5]{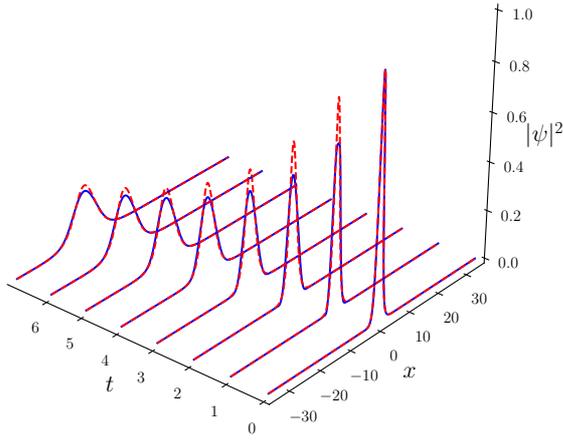}\label{subfig:FS_sech_good1}}
	\subfigure[]{\includegraphics[scale=0.5]{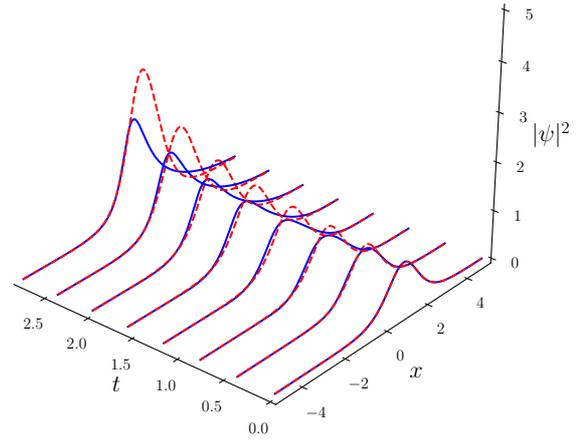}\label{subfig:FS_sech_bad1}}
	\subfigure[]{\includegraphics[scale=0.5]{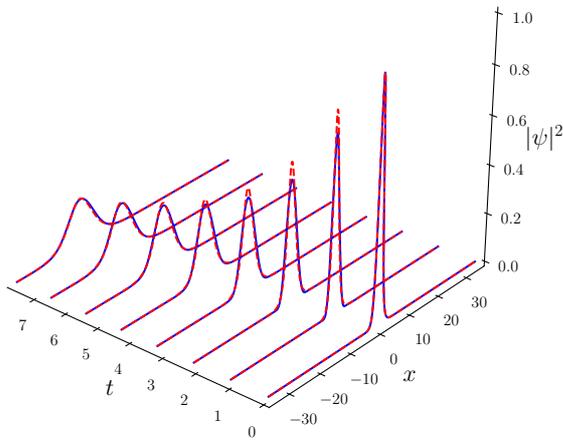}\label{subfig:DS_sech_good1}}
	\subfigure[]{\includegraphics[scale=0.5]{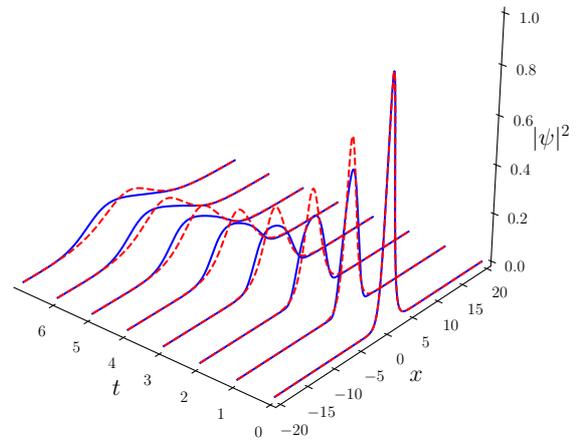}\label{subfig:DS_sech_bad1}}
	\caption{
The same as Fig.\ \ref{fig:gauss1} but for the $\sech$ ansatz \eqref{ansatz}.
		The parameter values for panels (a)-(d) are the same with those in Figs.\ \ref{subfig:FS_gauss_good1}, \ref{subfig:FS_gauss_bad1}, \ref{subfig:DS_gauss_good1}, and \ref{subfig:DS_gauss_bad1}, respectively.
			}
	\label{fig:FS_sech}
\end{figure}

\begin{figure}[htbp]
	\centering
	\subfigure[]{\includegraphics[scale=0.5]{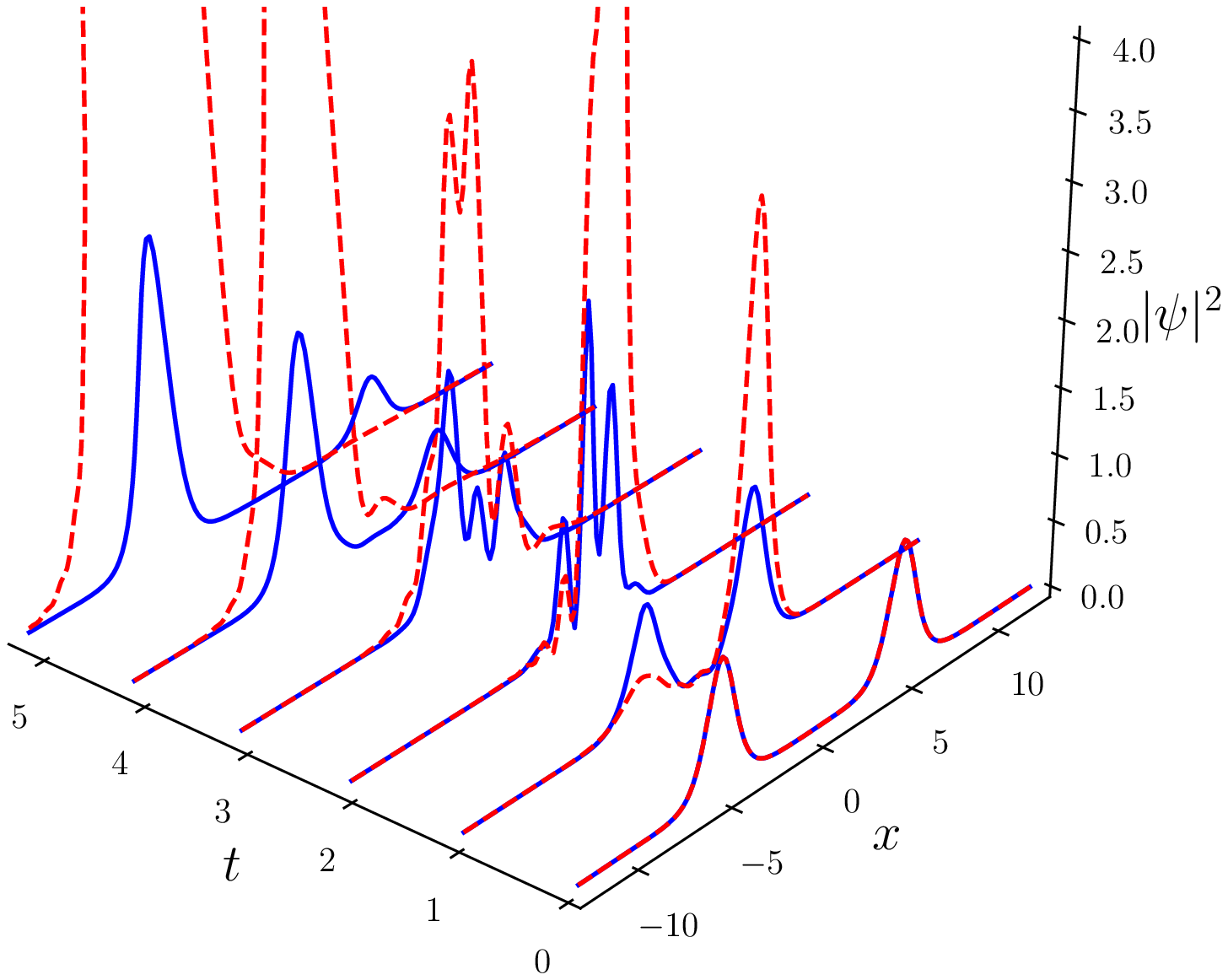}\label{subfig:FI_sech_good1}}
	\subfigure[]{\includegraphics[scale=0.5]{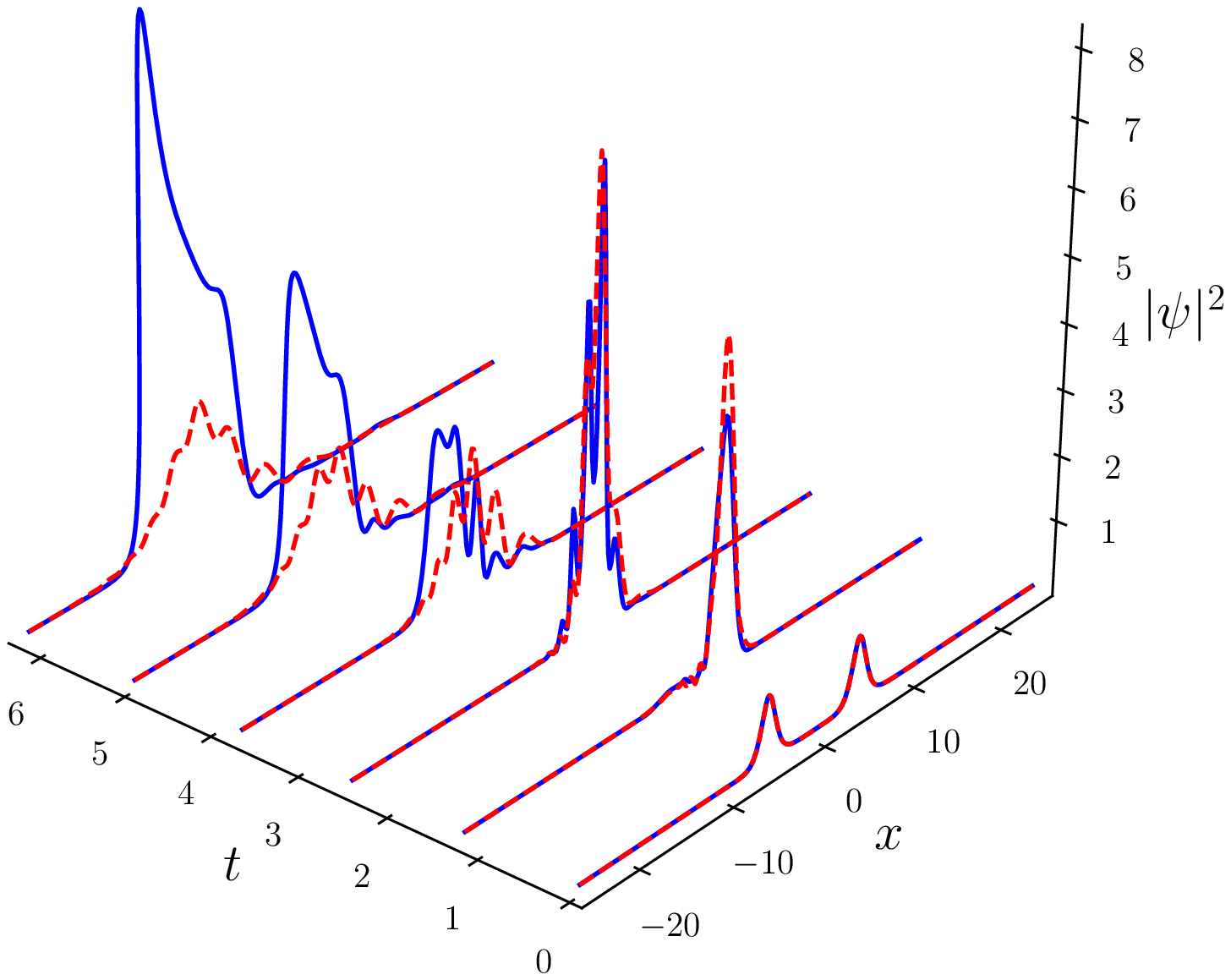}\label{subfig:FI_sech_bad1}}
\subfigure[]{\includegraphics[scale=0.5]{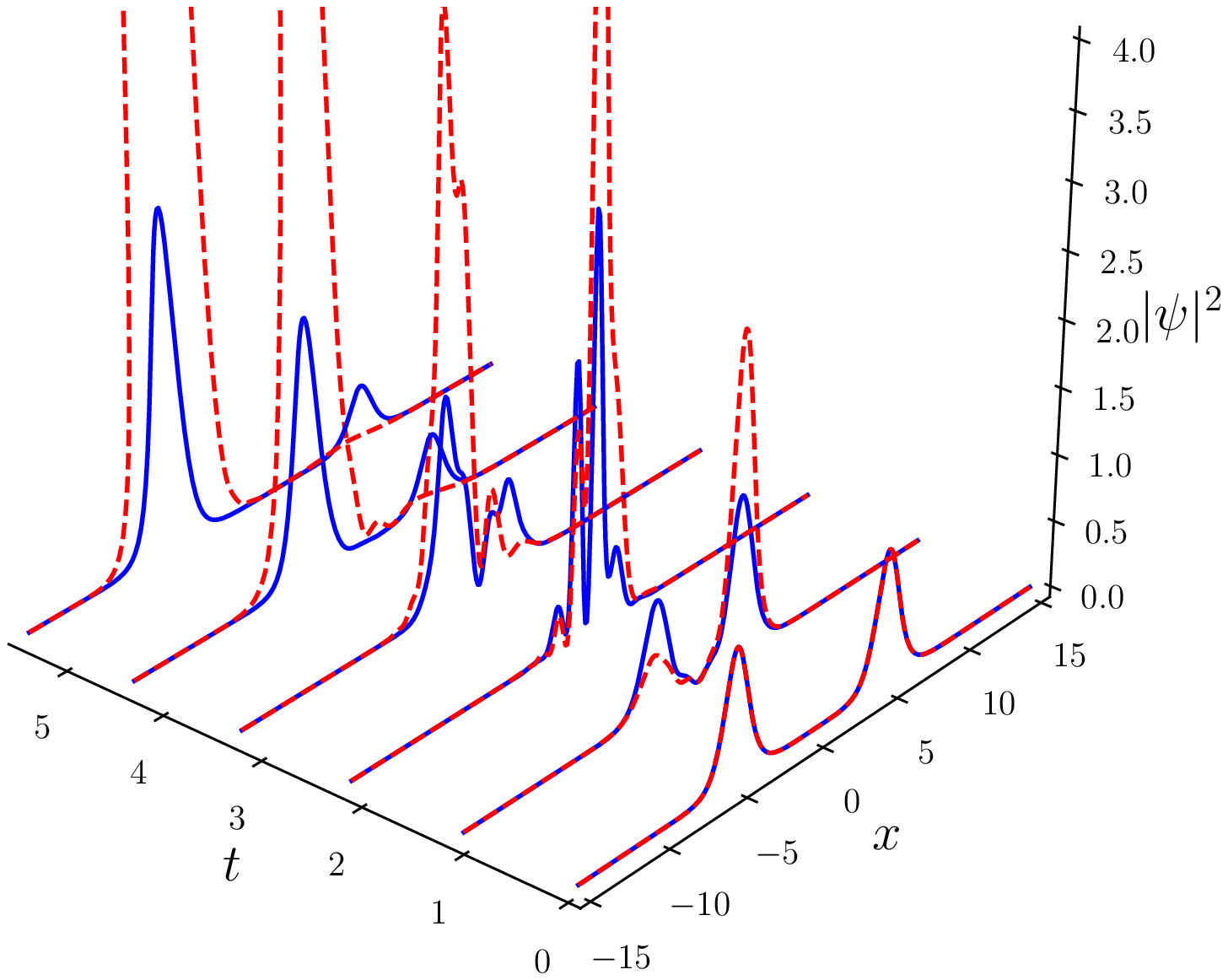}\label{subfig:DI_sech_good1}}
	\subfigure[]{\includegraphics[scale=0.5]{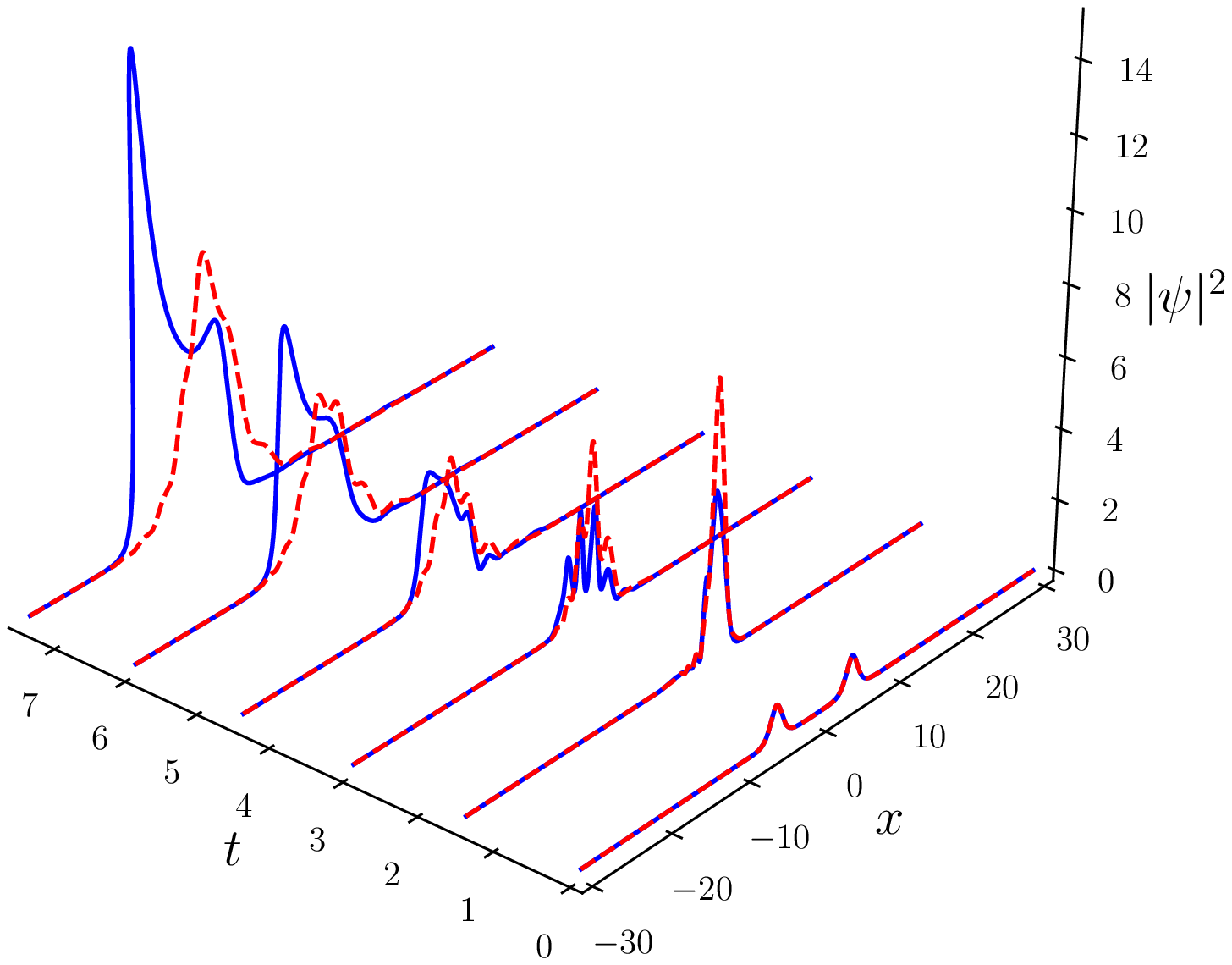}\label{subfig:DI_sech_bad1}}
	\caption{Soliton collisions in the focusing (panels (a) and (b)) and defocusing (panels (c) and (d)) case, comparing numerical and variational solutions for the $\sech$ ansatz \eqref{ans2}. 	The parameter values at $t=0$ are $A_1=A_2=1$, $B_1=0$, $C_1=C_2=1$, $D_1=D_2=0$, $E_1=-E_2=2$, $X_1=-X_2=5$, with (a) $B_2=0.1$, (b) $B_2=1$ , (c) $B_2=\pi+0.1$, and (d) $B_2=\pi+1$.
			}
	\label{fig:FI_sech}
\end{figure}

For the blow-up case in Fig.\ \ref{subfig:FS_gauss_bad1}, we know from \cite{ablo13} that the solution is likely to be in the form of a sech shape. It is therefore suggestive to use the same shape ansatz. Using \eqref{ansatz} as the initial condition, we plot in Fig.\ \ref{fig:FS_sech} the dynamics of the initial condition for the same parameter values as in Fig.\ \ref{fig:gauss1}.

For large incoming velocities, we still obtain good agreement between the numerics and the VA, see Figs.\ \ref{subfig:FS_sech_good1} and \ref{subfig:DS_sech_good1} for the focusing and defocusing case, respectively, where the soliton is spreading after passing the origin. A similar spreading is also seen for small incoming velocities in the case of defocusing nonlinearity as shown in Fig.\ \ref{subfig:DS_sech_bad1}. Figure \ref{subfig:FS_sech_bad1} shows that ansatz \eqref{ansatz} together with \eqref{ODE2} can capture the blow up, even though only qualitatively. However, if one takes the initial position $|E(0)|\ll|X(0)|\ll1$, we will obtain better agreement. The discrepancy is due to our ansatz \eqref{ansatz} that is far away from the actual blow-up solution provided in, e.g., \cite{ablo13}.

Finally, we have also simulated collisions of two bright solitons moving toward each others, see Fig.\ \ref{fig:FI_sech}. The initial conditions are taken as per expression \eqref{ans2}. We consider different initial parameters between the two solitons. The difference is introduced through the phases $B_j$. We do not present interactions of twin solitons because in that case the governing equation \eqref{nnls} will correspond to the local case, which has been considered for the first time in \cite{karp81}. 

Note that in contrast to the local NLS equation, nonlocal NLS equation  with %coupling between the solitons, note that 
defocusing nonlinearity will still allow for two bright solitons, provided that their phases differ by $\pi$. Therefore, in here we also consider defocusing cases with a phase difference between the solitons a little bit different from $\pi$. 

From Fig.\ \ref{fig:FI_sech} we obtain that in general solitons will blow up in time. When they are almost identical, i.e., Figs.\ \ref{subfig:FI_sech_good1} and \ref{subfig:DI_sech_good1}, they will pass each other at the origin. However, %following the collision, 
there is a continuous transfer of mass from one to the other. % while they travel away from the origin. 
Depending on the initial phases, one of them increases in amplitude while the other one vanishes. From the numerical simulations we obtain that the mass transfer occurs from a soliton with positive relative phase to that with a negative one. 

Such transfers can be explained from viewing the nonlocal NLS equation \eqref{nnls} as a Hamiltonian with the self-induced complex potential $V(x,t)=\sigma\psi(x,t)\psi^*(-x,t)$. Soliton with the increasing amplitude has Im$[V(x,t)]<0$, i.e., it experiences 'gain'. The other soliton depletes because it experiences 'loss', i.e., Im$[V(x,t)]>0$. Moreover, we have the conserved quantity \cite{ablo13}\begin{equation}Q=\int_{-\infty}^\infty\psi(x,t)\psi^*(-x,t)\,dx,\end{equation}called 'quasipower', that can be obtained from the power/mass of the NLS counterpart simply by replacing $\psi^*(x,t)$ with $\psi^*(-x,t)$. As some part of $\psi(x,t)$ increases in time, to keep $Q$ conserved, the other part of $\psi(x,t)$ will have to decrease. All these mechanisms create the dynamic effect of a continuous mass transfer between the solitons.

We also consider the situation when initially solitons are not quite identical, see Figs.\ \ref{subfig:FI_sech_bad1} and \ref{subfig:DI_sech_bad1}. In this case the solitons collide and then keep overlapping, instead of passing each other, with the tails that also grow in time. Both soliton peaks blow up later on (not shown in the figures). 

In all the cases, we see that our VAs describe the numerics well prior to soliton collisions. After the two solitons meet at the origin, VAs only capture their qualitative dynamics, such as blow up in Figs.\ \ref{subfig:FI_sech_good1} and \ref{subfig:DI_sech_good1} and merger in Figs.\ \ref{subfig:FI_sech_bad1} and \ref{subfig:DI_sech_bad1}.

\section{Conclusion}
\label{sec6}

In this paper, we considered the integrable nonlocal nonlinear Schr\"odinger equation proposed by Ablowitz and Musslimani. We have derived a collective coordinate approach based on variational methods to study dynamics (including collisions) of moving solitons. Through comparisons with numerical computations, we have examined its applicability in the system. We obtained that VAs are generally good in describing single soliton dynamics. 

For collisions of two solitons, they capture the dynamics quite well before collisions at the origin, while afterwards they only provide qualitative comparisons. Our current work implies that VAs should be well applicable to describe soliton dynamics in the evolution equations \cite{ablo04}
\begin{align}
&iq_t+\frac12q_{xx}+\sigma q^2r = 0,\\
&ir_t-\frac12r_{xx}-\sigma r^2q = 0,
\end{align}
which give the local NLS equation  if $r=\pm q^*(x,t)$ and the nonlocal NLS equation  \eqref{nnls} if $r=\pm q^*(-x,t)$. However, the general case when $r(x,t)$ is not explicitly related to $q(x,t)$ through a symmetry has not been considered yet. %This is addressed for future work. 
It will be particularly relevant to extend our study to this general system of equations. In that case, ensembles of $N$-solitons, as the work of Gerdjikov et al.\ \cite{gerd96a,gerd96} for the classical NLS equation, will be interesting. 

For future works, it will be particularly important to seek for further approximation ansatzs to make the VA better approach the numerical results. Such an ansatz would need to be close enough to the actual solutions we consider. However, balancing between accurate approximations and simple computations is unfortunately not an easy task.  Additionally, VAs for dark solitons in the defocusing nonlocal NLS equation  are also interesting to be developed. 

\section*{Acknowledgments}
R.R (Grant Ref. No: S-5405/LPDP.3/2015) and R.K (Grant Ref. No: S-34/LPDP.3/2017) gratefully acknowledge financial support from Lembaga Pengelolaan Dana Pendidikan (Indonesia Endowment Fund for Education). The authors thank the three reviewers for their comments that improved the quality of the paper. \\

The authors contribute equally to the manuscript. 

%\section*{References}

\end{document}